\documentclass[runningheads]{llncs}

\usepackage{booktabs} %
\usepackage{graphicx}
\graphicspath{{./figures/}}
\usepackage{subcaption}

\usepackage{tikz}
\usepackage{tikzscale}
\usepackage{adjustbox}
\usepackage{pgfplots,pgfplotstable}
\usepackage{smartdiagram}
\usetikzlibrary{positioning}
\usetikzlibrary{pgfplots.statistics,shapes,arrows,backgrounds}
\pgfplotsset{compat=newest}
\usepackage{xspace}
\usepackage{hyperref}

\newcommand{\dyntest}{\textsc{DynTest}\xspace}

\newcommand{\evalproject}{IPS}

\begin{document}

\title{Constraint-Guided Test Execution Scheduling: An Experience Report at ABB Robotics}

\titlerunning{Constraint-Guided Test Execution Scheduling}
\author{Arnaud Gotlieb\inst{1}\and
Morten Mossige\inst{2} \and
Helge Spieker\inst{1}
}
\institute{Simula Research Laboratory\footnote{List of authors is given in alphabetical order}, Kristian Augusts gate 23, 0164 Oslo, Norway
\email{\{arnaud,helge\}@simula.no}\\
\and
ABB Robotics, Bryne, Norway \email{morten.mossige@uis.no}
}

\maketitle

\begin{abstract}
Automated test execution scheduling is crucial in modern software development environments, where components are frequently updated with changes that impact their integration with hardware systems.  
Building test schedules, which focus on the right tests and make optimal use of the available resources, both time and hardware, under consideration of vast requirements on the selection of test cases and their assignment to certain test execution machines, is a complex optimization task.
Manual solutions are time-consuming and often error-prone. Furthermore, when software and hardware components and test scripts are frequently added, removed or updated, static test execution scheduling is no longer feasible and the motivation for automation taking care of dynamic changes grows. 
Since 2012, our work has focused on transferring technology based on constraint programming for automating the testing of industrial robotic systems at ABB Robotics. After having successfully transferred constraint satisfaction models dedicated to test case generation, we present the results of a project called \dyntest{} whose goal is to automate the scheduling of test execution from a large test repository, on distinct industrial robots. This paper reports on our experience and lessons learned for successfully transferring constraint-based optimization models for test execution scheduling at ABB Robotics.   
Our experience underlines the benefits of a close collaboration between industry and academia for both parties.
\end{abstract}

\section{Introduction} \label{sec:introduction}
Continuous integration (CI) has been adopted by many companies all around the world in order to ensure better end-user product quality~\cite{Klo22}. As part of CI, automated testing is crucial to get quicker feedback on the detected defects or regressions of a system under test. When a complete industrial system is tested under CI, a challenge arises if it relies on hardware and software components, because they can hardly be tested in isolation. Besides, additional challenges include the requirement to generate tests with environmental hazards, the combinatorial explosion of the number of potential test cases due to parameter interactions, the automation of test execution scheduling which ensures proper coverage and diversity of test cases and agents.

This paper reports on our experience in deploying a constraint-guided test execution scheduling method as part of a CI process at ABB Robotics. By co-developing an automated testing process named \dyntest{} through an industrial-academic partnership, the authors have explored the transfer of advanced constraint programming\footnote{Constraint Programming is a declarative programming framework which uses relations among logical variables and search procedures to find solutions of combinatorial problems \cite{Rossi2006}.} models composed of global constraints and rotational diversity \cite{SGM19} in a highly automated industrial testing process \cite{Gotlieb2021}. Since 2012, multiple models for test case generation \cite{MGM14a,MGM15} and selection, test prioritization \cite{SGM17} and eventually test execution scheduling \cite{MGS17} have been explored, evaluated and transferred. Our experience underlines the benefits of a close collaboration between industry and academia for both parties in the area of automated testing.

\section{Test Execution Scheduling at ABB Robotics} \label{sec:problem}
\vspace*{-0.9cm}
\begin{figure*}[th]
  \centering
     \includegraphics[width=10cm]{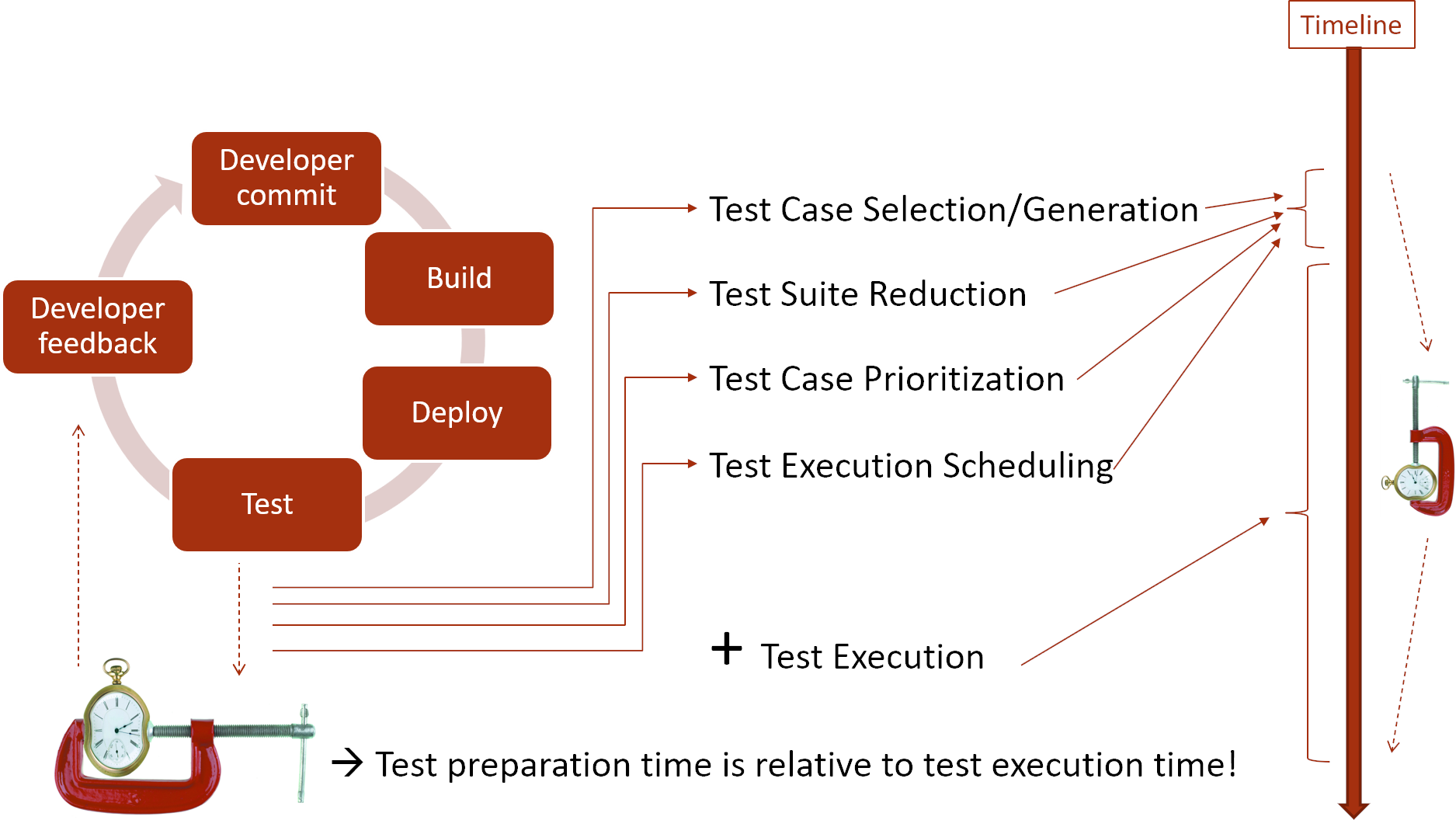}
  \caption{Overview of the CI cycle and the challenge related to time management}
  \label{fig:CIcycle}
\end{figure*}
\vspace{-0.5cm}
ABB Robotics is an industrial robot supplier and manufacturer company operating in more than 50 countries around the world. A key objective of the company is to deliver high-quality products (thus involving an increased focus on testing robots for reliability and performance) for the benefice of its customers. Initially, robot testing was done mostly manually and using human-eyes visual control for checking the results of hand-crafted tests. 
This restricted the possible testing time to human-worked hours of test engineers (besides long-running tests, which could use nighttime and weekends) and did not use available robot to its full test capability. To reduce the time-to-market of new products and also improve the quality of these products, the testing process had to be much more automated.
To start with, the test automation process had to be placed within a Continuous Integration (CI) process. 

As shown in Fig.~\ref{fig:CIcycle}, a typical CI cycle includes software developer commit actions which automatically trigger build, deploy and test activities. The test results are then passed back to the developers to provide them with feedback. Typically, the test activity includes the following five steps:
\begin{enumerate}
    \item {\bf Test Case Selection and Generation:} Tests are either extracted from an existing repository or automatically generated from specific requirements;
    \item {\bf Test Suite Reduction:} Test suites that achieve a given objective (e.g., full requirement coverage) are pruned to eliminate spurious test cases;
    \item {\bf Test Case Prioritization:} Tests are ordered to provide a quick feedback by using either pre-determined or dynamically-computed priority values;
    \item {\bf Test Execution Scheduling:} Test plans are distributed on different robots, in a specific order according to a pre-computed test schedule;
    \item {\bf Test Execution:} Tests are then eventually executed according to the specified schedule, in order to identify defects in the system under test. This activity is clearly the most demanding as it requires launching the system with the test cases selected and prioritized in the previous steps.
\end{enumerate} 
It is worth noting that, in CI, controlling the test preparation time (i.e., the four first steps) with respect to the test execution time (i.e., the fifth step) is crucial. Knowing that the overall time-line allocated to test activities has to be bounded, we have to keep as much time as possible for test execution. Of course, an optimized test schedule (computed during test preparation) can lead to better test execution, but it makes no sense to spend too much time in the computation of a schedule if it reduces too much the time available for test execution. As shown in Fig.~\ref{fig:CIcycle}, finding the right trade-off is part of the testing challenges faced at ABB.

In ABB's context, a {\it test case} aims at verifying a robotized task, which is performed by a robot under the observation of some sensors. 
A test can either fail or succeed; it fails when observations reveal a misfunction, and it succeeds when no misfunction is observed. 
A test case is associated with some metadata, consisting of its average duration, previous execution times, results, and targeted robots.
Each test case execution is non-preemptive, that is, it cannot be interrupted by another test or transferred to another robot during execution.
All test cases are independent, without any dependency on the order in which they are executed. 
Still, they can be ordered by using their static priority, which is decided by the test engineers, and their dynamic priority, which is based on a combination of their effectiveness to reveal defects in earlier CI cycles and the time since their last execution.
Test cases have furthermore hardware requirements, meaning that they can only be executed on certain robots.

Test cases are executed by \textit{test agents}, which are software components that capture the various schedules computed for each CI cycle. Each test agent has a limited amount of time available per cycle and a set of compatible test cases, which it can execute.
Computing a test schedule requires to vary the assignment between test cases and test agents between cycles to achieve a full coverage of all possible combinations between tests and hardware over time. 
Fulfilling this objective balances the confidence in the stability of certain features on different hardware, while giving room for executing many test cases and not executing the same tests multiple times during a cycle.

\section{Automated Testing Process}\label{sec:approach}
Here, we present the approach to automate the testing process within the CI environment.
The testing process is sequential with distributed components, orchestrated by a central test controller. Starting with {\it data initialization and acquisition} (Sec.~\ref{sec:data}), the process computes the priorities over test cases (Sec.~\ref{sec:prio}) and the test schedules (Sec.~\ref{sec:schedule}). Test execution is performed by distributing test plans to each robot (Sec.~\ref{sec:exec}) and eventually test execution reporting takes place (Sec.~\ref{sec:report}). 
The central test controller, referred to as \dyntest{}, manages the process, acquires and distributes the necessary data from other sources, and provides the interface towards the automated testing process. Other components include a module for test case prioritization, selection, and scheduling, and a module for controlling test agents executing the test cases.

\subsection{Data Initialization and Acquisition}\label{sec:data}
The process is set up with available test cases and agents. Some test cases and agents are filtered out to exclude scripts and robots under maintenance or having incompatible hardware requirements.
For the remaining test cases, historical meta-data is extracted from the central data repository. This data includes the most recent test execution results, their runtime, and previous test agents they were executed on, etc.

\subsection{Test Case Prioritization}\label{sec:prio}
The prioritization step is initially designed with a simple approach, to ease the setup of automation and definition of assigned priorities during integration by the test engineers.
The process iterates through all executable test cases and assigns each a priority, which is a weighted sum of the time since the last run, the test case duration, and the most recent last results
The weights and number of considered historical test results are manually chosen during integration, but that process could be replaced by a self-adaptive method in the future. %

\subsection{Selection and Scheduling}\label{sec:schedule}
\begin{figure}[t]
  \centering
\includegraphics[width=8cm]{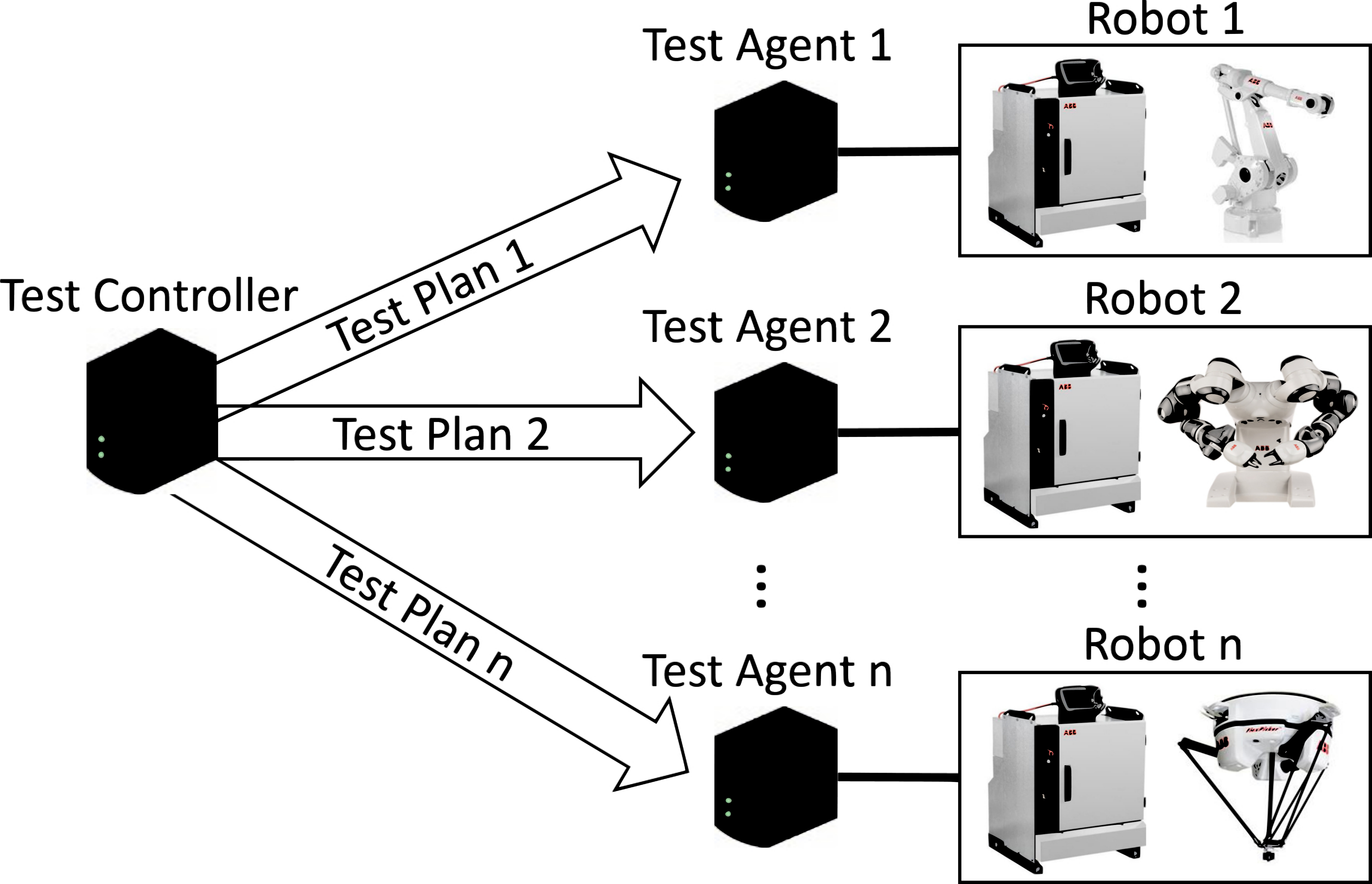}
\vspace*{-0.2cm}
  \caption{The test controller distributes individual test plans to each test agent, which controls, in turn, one robot and records log information and test outcomes.
  \label{fig:distribution}}  
  \vspace*{-0.6cm}
\end{figure}
Selection and scheduling focus on taking the test cases with the highest priorities and distributing them to the test agents until all time available for testing is used. Although test case selection and scheduling are often regarded as two separate tasks in the literature, in practice, we closely integrate both steps. Selection means to take those test cases from the set of prioritized test cases, which are most desirable to execute. Because the execution of test cases is constrained, this selection has to consider which subset of test cases can actually be executed and at the same time maximizes the available resources (as we want to avoid idle times). The selection and scheduling step receives a set of prioritized test cases and a set of available test agents as inputs. During this step, \dyntest{}{} creates an execution schedule, where each test case is assigned to one test agent for execution while preferring to assign high-priority test cases over low-priority test cases. During selection, test cases which are marked as obligatory to be run, are always included in the final schedule, regardless of their calculated priority.

We now approach this scheduling task by using Constraint Programming (CP), even if, in its initial version,  
only a simple greedy first-fill algorithm was used. This heuristic algorithm's first ordered the test cases by descending priority. Then, successively for each test agent, the test case with the highest priority was assigned to the test agent until the maximum time limit was reached. However, we quickly discovered that this too-simplistic approach was not suitable to ensure sufficient diversity in the selection of test cases and agents. We then developed a refined model based on CP.
CP is a paradigm in which a problem is not modelled as a sequence of steps to achieve a desired solution, i.e., an algorithm, but relations between variables are described to formulate properties of a desired solution~(see \cite{Rossi2006}).
CP and its associated optimization methods are efficient and well-performing techniques for modelling strictly constrained problems, such as planning and scheduling problems~\cite{Bartak2010}.
Using CP for scheduling enables precise control over the execution time and the trade-offs made between time looking for a solution and the solution's quality.
We replaced the initial scheduling method with a dedicated constraint optimization model which further optimizes the schedules by ensuring that the assignment between test cases and test agents changes between test cycles. We called this process rotational diversity and used global constraints to develop it. Full details on this constraint model are available in \cite{SGM19}.

\subsection{Distribution and Execution}\label{sec:exec}
\begin{figure}[t] 
  \includegraphics[width=\textwidth]{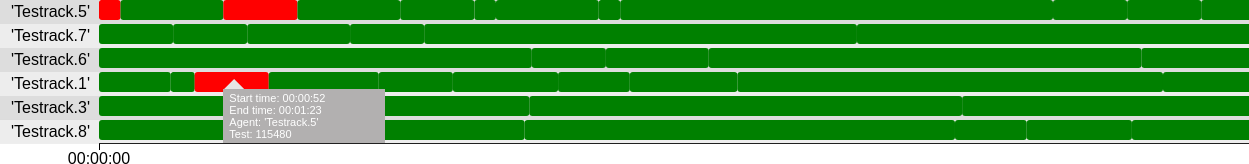}
    \caption{Visualization of a test schedule with interactive access to test results.}
    \label{fig:schedule}
\end{figure}
Once the test schedule is created, \dyntest{} transforms it into separate, individual test plans and sends them to the corresponding test agent. An example test plan is shown in Fig.~\ref{fig:schedule}. Each test agent executes all assigned test cases independently, as there is no interdependency between test cases and robots.
The test agent records all test results and log files from the test cases and returns them to the test controller.

\subsection{Reporting}\label{sec:report}
\begin{figure}[t]
  \begin{subfigure}[T]{0.5\textwidth}
  \includegraphics[width=\textwidth]{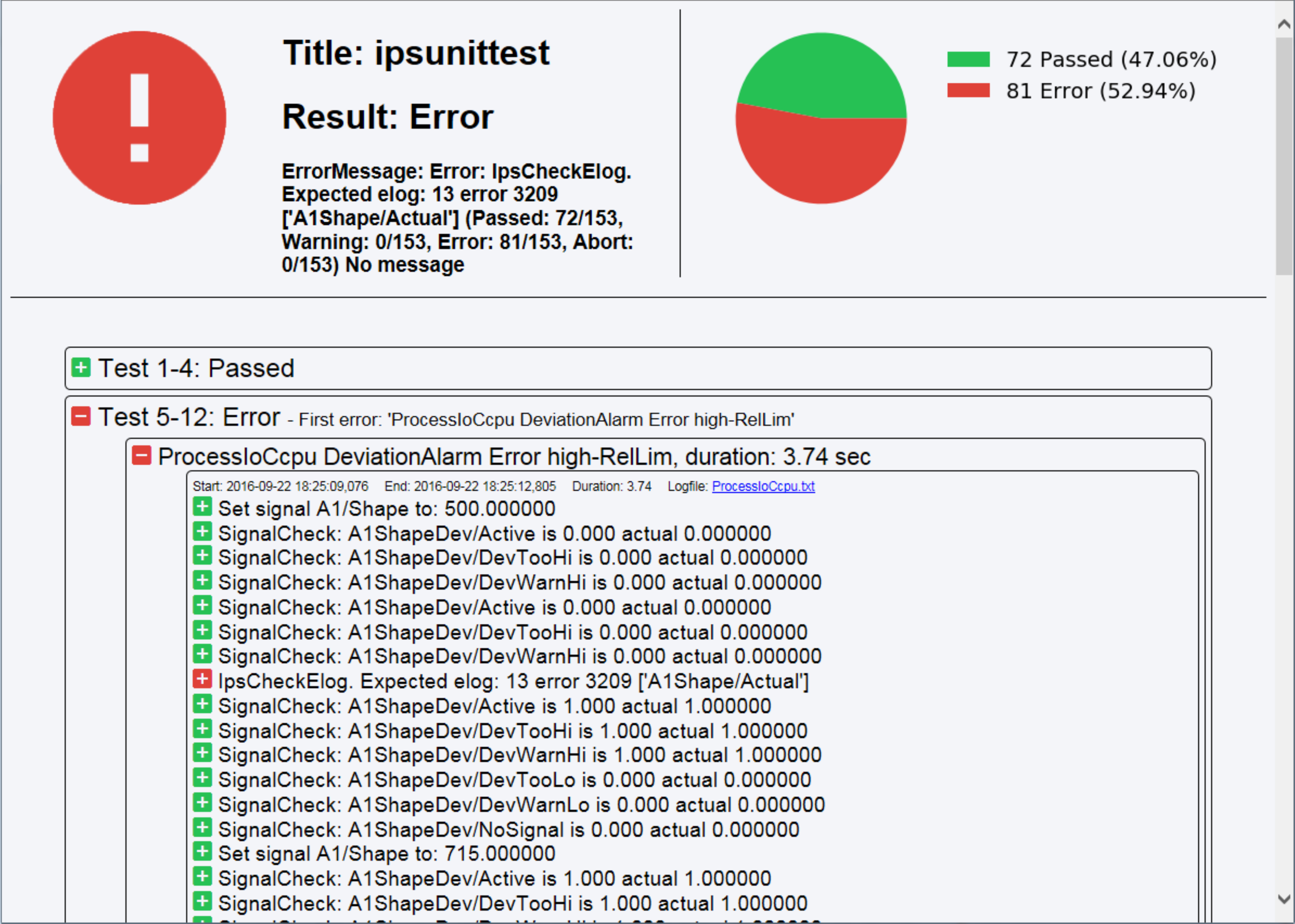}
   \end{subfigure}\hfill%
    \begin{subfigure}[T]{0.5\textwidth}
  \resizebox{0.85\textwidth}{!}{\includegraphics{utilization.pgf}}
  \resizebox{0.85\textwidth}{!}{\includegraphics{priority.pgf}}
   \label{fig:utilization}
   \end{subfigure}
   \caption{Visualization of a test report for the IPS project and distribution of resource usage and test case priority among CI cycles for a one-month period.}
   \vspace*{-0.6cm}
   \label{fig:testreport}
\end{figure}

Reporting aims to communicate the test execution results back to the developer for failure analysis. An example of a test report is shown in Fig.~\ref{fig:testreport}. The report summarizes the results of a test cycle, allows us to navigate into lower levels of the test hierarchy and access specific details of single test executions. This hierarchical structure makes the report accessible to different user groups and is the first step of debugging and failure analysis. Another goal of the reporting step is to gather and visualize information about the testing process itself. A visual report of the scheduling outcome is created as a means for individual run analysis and communication (see Fig.~\ref{fig:testreport}). It is built on web technologies and enables interactive exploration, including access to test case information and results of recent executions. 
Exhaustive reporting and data collection enables better long-term evaluation of the system's behavior as well as impact evaluation onto software development, which is an important aspect for tuning the process in the future.

Besides the reporting of individual test results, monitoring the overall behaviour of \dyntest{} is performed.
Fig.~\ref{fig:testreport} shows examples of two such monitoring metrics. The resource utilization monitors how efficiently the available resources are filled by the test case scheduling algorithm, here most plans should show a high utilization of close to 100\% to make the best use of the available resources. The distribution of test case priorities shows the variation in relevance of test cases. Here, there is a large block of highly important test cases with high priority but also chunks with low priority as well as average priority, indicating a good overall balance of priorities.

\section{Empirical Evaluation} \label{sec:evaluation}

After a development phase where the integration of all steps of the automated testing process was realized, we performed a one-month empirical evaluation of an existing subsystem called IPS (Integrated Painting Systems).
Even though an exhaustive quantitative evaluation of the testing process is difficult as it substantially impacts the working processes, we drew some conclusions on the process by examining the schedules created by \dyntest{}. 
For the evaluation, we considered $87$ CI cycles of \evalproject{}.
As stated above, Fig.~\ref{fig:testreport} reports on the resources utilization and test case priority of the test schedules of \evalproject{}. Each schedule achieves a resource utilization of at least 91\,\% with the majority having a utilization of 99\,\% meaning that the available time for testing is used extensively.
An overall utilization of 100\,\% is not achievable for two reasons.
First, the total duration of test case execution is not guaranteed to sum up to the total available time.
Second, during scheduling, the focus is on assigning highly prioritized test cases and then filling in the available time with the most important test cases instead of maximizing the time usage.
Regarding test case priority, Fig.~\ref{fig:testreport} shows that the test cases are spread among the spectrum of possible priorities, with two noticeable clusters at the lower and upper bound of the spectrum. %
Having a similar number of high- and low-priority test cases stems from the fact, that high-priority test cases, once they have passed their last execution, tend to receive a low priority during the next cycle. This behavior distinguishes from test cases which have not failed during the observed period. After having not been executed for a while, the priority grows again and these test cases become likelier to be executed again.

\section{Lessons Learned}\label{sec:lessonslearned}
We report on three lessons learned while developing test automation process.\\
\noindent\textbf{Automated test scheduling through CI is crucial to improve robot software/hardware quality.} Automated testing through CI allows us to detect at an early stage hardware/software defects on robots and avoid the propagation of failure at customer sites. It also reveals regression issues when the specification of a new product is not yet finalized. This approach significantly improves the overall product quality;\\
\noindent\textbf{Incremental co-development is relevant when complex constraint optimization models have to be developed.} We co-developed a test execution scheduling component as part of \dyntest{}.   
Starting from a simple version (based on an inefficient greedy-based scheduling approach), we developed a complex constraint optimization model based on global constraints and rotational diversity incrementally. 
This approach was key to fostering the adoption and maintenance of this complex model by people who do not necessarily have the expertise to maintain advanced constraint models;\\
\noindent\textbf{Industry-academic co-development.} The outcomes of this co-development were beneficial for both sides. 
On one hand, ABB Robotics benefited from the academic expertise in constraint-based scheduling, which was required to develop test execution scheduling models. 
On the other hand, scientists took advantage of the industrial experience of the test engineers in the test automation processes, to publish advanced research results with empirical results. 
Finally, thanks to this co-development, the transferability of the method was easier.

\section{Conclusion} \label{sec:futurework}
This paper reports on an experience to transfer constraint-based models for automated test execution scheduling at ABB Robotics. 
In this work, advanced constraint-based scheduling models using global constraints and rotational diversity were developed and empirically evaluated, and industrialized as part of a complete CI process. 
Further work includes refinement in the description of test cases to handle specific globally-shared external equipment.

\bibliographystyle{splncs04}
\bibliography{refs}

\end{document}